\documentclass[twocolumn,prb]{revtex4-1}
\usepackage[english]{babel}

\usepackage{graphicx}
\usepackage{amsmath}
\usepackage{amsfonts}
\usepackage{amssymb}
\usepackage{bm,bbm}
\usepackage{color}

\usepackage{hyperref}

\newcommand{\sx}[1]{\sigma^{\rm x}_{#1}}
\newcommand{\sy}[1]{\sigma^{\rm y}_{#1}}
\newcommand{\sz}[1]{\sigma^{\rm z}_{#1}}
\def\ii{{\rm i}}

\newcommand{\tr}{\mathrm{tr}}

\def\tit#1{{\em #1},}
\def\etal#1{#1}

\usepackage{booktabs}
\usepackage{multirow}
\AtBeginDocument{
\heavyrulewidth=.08em
\lightrulewidth=.05em
\cmidrulewidth=.03em
\belowrulesep=.65ex
\belowbottomsep=0pt
\aboverulesep=.4ex
\abovetopsep=0pt
\cmidrulesep=\doublerulesep
\cmidrulekern=.5em
\defaultaddspace=.5em
}

\begin{document}

%\title{Modified Matthiessen's rule: breaking anomalous transport to diffusion by dilute scattering}
%\title{Modified Matthiessen's rule: breaking of the additivity of scattering rates due to anomalous transport}
%\title{Less is more: breaking the additivity of scattering rates by anomalous transport}
\title{Less is more: more scattering leading to less resistance}
%\title{Modified Matthiessen's rule: more scattering leading to less resistance}

\author{Marko \v Znidari\v c}
\affiliation{Department of Physics, Faculty of Mathematics and Physics, University of Ljubljana, Jadranska 19, SI-1000 Ljubljana, Slovenia}

\date{\today}

\begin{abstract}
We study the breaking of integrability by a finite density of dilute impurities, specifically the emerging diffusive transport. Provided the distance between impurities (localized perturbations) is large, one would expect that the scattering rates are additive, and therefore, the resistivity is proportional to the number of impurities (the so-called Matthiessen's rule). We show that this is, in general, not the case. If transport is anomalous in the original integrable system without impurities, the diffusion constant in the non-integrable system at low impurity density gets a nontrivial power-law dependence on the impurity density, with the power being determined by the dynamical scaling exponent of anomalous transport. We also find a regime at high impurity density in which, counterintuitively, adding more impurities to an already diffusive system increases transport rather than decreases it.
\end{abstract}

\maketitle

\section{Introduction}

A kinetic Drude model of transport~\cite{Kittel,Mermin} has been proved again and again to be a rather useful picture despite its ``cartoonish'' simplicity (or, perhaps precisely because of it). The electric current $J$ (or a current of any other conserved quantity) is given by $J=nev$, where $en=eN/V$ is the density of the conserved charge and $v$ is a ``characteristic velocity''. The Drude model explains finite conductivity $\sigma$ as being due to rare scattering events that perturb an otherwise ballistic motion of electrons. With $\tau$ being the scattering, i.e., the relaxation time, the velocity an electron initially at rest will reach in time $\tau$ under a uniform acceleration $eE/m$ is $v=eE\tau/m$, resulting in $J=\frac{ne^2\tau}{m}E$. In other words, in the Drude model the conductivity depends on the single parameter $\tau$ as $\sigma=\frac{ne^2\tau}{m}$. While the above is not much more than a dimensional analysis, the question one has to answer for a particular situation is what determines $v$, or, equivalently $\tau$, e.g., does it actually correspond to a velocity of any real excitation. Nevertheless, it does highlight the crucial role played by a characteristic time (velocity) in such a ballistic picture.

Now imagine a material in which several different types of scatterings can occur, each characterized by its own scattering time $\tau_k$. Provided scattering events are separated one would argue that they are independent and therefore one can just add the individual scattering rates to obtain the total rate $1/\tau$ as
\begin{equation}
\frac{1}{\tau}=\sum_k \frac{1}{\tau_k}.
\label{eq:add}
\end{equation}
The above additivity principle, being equivalent to saying that in Fermi's golden rule we have to add probabilities instead of amplitudes, means that if we have $K$ impurities each causing scattering with rate $1/\tau_0$, one will have $\tau=\tau_0/K$. Therefore, the conductivity will scale as
\begin{equation}
\sigma \propto \frac{1}{K},
\label{eq:lin}
\end{equation}
or, equivalently, resistivity will be linearly proportional to the number of impurities $K$.

Such Matthiessen's rule~\cite{matthiessen,Kittel,Mermin,Ziman}, Eqs.~(\ref{eq:add},\ref{eq:lin}), can be for instance observed in metals at low temperatures where the phonon scattering is negligible and impurities dominate, resulting in the resistivity being proportional to the concentration of impurities. Matthiessen's rule is rather natural -- it predicts that if one adds twice as many impurities to a clean metal the resistivity will be twice as large. We will show that this textbook fact is in fact not correct if the clean material without impurities is not ballistic. In such a case the rule has to be modified to
\begin{equation}
\sigma \propto \frac{1}{K^{2-z}},
\end{equation}
where $z$ is the dynamical (transport) exponent of the clean system (e.g., a ballistic system has $z=1$, diffusive $z=2$, subdiffusive $z>2$). Therefore, resistivity is in general not proportional to $K$. Interestingly, for $z>2$ one can even have a regime where adding more impurities will actually increase conductivity. 

We note that while violations of Matthiessen's rule have been observed before~\cite{Bass,Klein01}, e.g., due to non-isotropic scattering, they are rather small. Here we present a mechanism for a complete and conceptually new breakdown of the rule. The idea was already put forward recently in Ref.~\onlinecite{prl20} where it was demonstrated for the isotropic Heisenberg model at high temperature which is superdiffusive with $z=3/2$. In the present work we shall verify the modified rule for a continuous set of anomalous transport coefficients $z$.
 
The type of model that we address is sketched in Fig.~\ref{fig:shema}, the main result that we verify is in Eq.~(\ref{eq:mod}), with the supporting data in Fig.~\ref{fig:D}.

\section{Fibonacci model}

\begin{figure}[t!]
  \centerline{\includegraphics[width=3.3in]{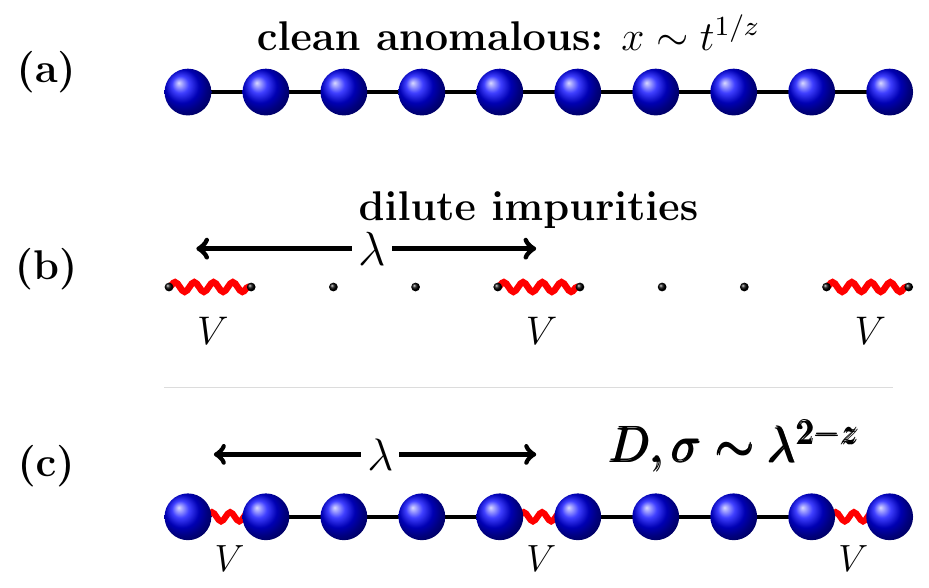}}
  \caption{We study transport in the system shown in (c) which has dilute integrability-breaking impurities (red wiggles). It is obtained by taking an integrable system with anomalous transport (a), and adding interaction at every $\lambda$-th bond (b). As shown in (c), this results in diffusive transport with the diffusion constant $D$ and the conductivity $\sigma$ scaling as $\lambda^{2-z}$ with the distance $\lambda$ between impurities (the additivity-based Matthiessen's rule predicts $D,\sigma \sim \lambda$).}
\label{fig:shema}
\end{figure}

We would like to study transport in a model with anomalous transport to which local perturbations (impurities) are added. We are in particular interested in the limit where the distance $\lambda$ between impurities is large, see Fig.~\ref{fig:shema}. While the limit $\lambda \gg 1$ represents a rather special type of perturbation, and this will allow for simple theory, it is very much the relevant limit for high-purity materials. For instance, in cuprates one has studied~\cite{Kawamata08,Hlubek11,Hlubek12} the influence of diagonal or off-diagonal disorder with concentrations $1/\lambda \approx 10^{-2}-10^{-4}$ on heat conductivity. We remark that we will be interested in the thermodynamic limit (TDL) while keeping $\lambda$ fixed, i.e., a finite density of impurities that will always result in normal diffusive (Ohmic) transport as opposed to for instance the case of a single impurity~\cite{Lea,speck,Brenes} which, while it changes an integrable system to a chaotic one according to standard criteria, it does not modify the system's transport~\cite{Brenes}. We also don't focus on details of how the broken integrability leads to finite diffusion constant, see e.g. Refs~\onlinecite{rosch00,jung06,huang13,robin14,vadim18,pnas18,comment21,diff,Doyon,Vasseur,Friedman20}.

We study a one-dimensional interacting Fibonacci model. Taking a one-dimensional spin-$1/2$ chain will allow for an efficient numerical assessment of spin transport at an infinite temperature. The Fibonacci model, which in its non-interacting version with $V_j\equiv 0$ will serve as our unperturbed clean model, is described by the Hamiltonian
\begin{equation}
  H=\sum_{j=1}^{L-1} \sigma_j^{\rm x}\sigma_{j+1}^{\rm x}+\sigma_j^{\rm y}\sigma_{j+1}^{\rm y}+V_j \sigma_j^{\rm z}\sigma_{j+1}^{\rm z}+\sum_{j=1}^L h_j\, \sigma_j^{\rm z}.
\label{eq:H}
\end{equation}
The on-site fields are given by the Fibonacci potential of amplitude $h$, $h_j=h\{2f(\beta j)-1\}$, where $\beta=(\sqrt{5}-1)/2$ and $f(x)=[x+\beta]-[x]$ with $[x]$ being an integer part of $x$. For instance, the beginning of the sequence is $h_j=h(+1,-1,+1,+1,-1,\ldots)$. Because $f(x)$ is periodic and $\beta$ is irrational the on-site potential is quasiperiodic. In the non-interacting model in which all interactions $V_j$ are zero this allows for non-trivial transport properties that are intermediate between the ballistic transport one would have for a periodic potential and localization for random $h_j$. Namely, the non-interacting Fibonacci model is critical~\cite{kohmoto83,ostlund83,kitaev86,sutherland87} for any $h$ and displays a continuously varying anomalous transport~\cite{HiramotoAbe}, going from ballistic ($z=1$) at $h=0$ to localized ($z=\infty$) in the limit $h \to \infty$. Due to its interesting physical and mathematical~\cite{damanik} properties the non-interacting Fibonacci model has been much studied~\cite{kohmoto83,ostlund83,kitaev86,sutherland87,HiramotoAbe}, including transport~\cite{vidal,lacerda21}, and has been realized in experiments~\cite{akkermans14,strkalj20}. 

The interacting Fibonacci model is less understood~\cite{vidal,Alet18,fibo19,goold21,doggen21}. We shall focus on a particular case of dilute interactions, that is, we will have nonzero interaction equal to $V_j=1$ at every $\lambda$-th site (see Fig.~\ref{fig:shema}). For such perturbation transport will be always diffusive in the TDL; as long as one has average $\lambda \gg 1$ none of our results, like Eq.~(\ref{eq:mod}), should depend on the precise form of a localized perturbation and the fact that perturbed sites are exactly $\lambda$ sites apart. While we shall use spin language and calculate the spin diffusion constant $D$, one could equivalently use the Jordan-Wigner transformation and use the language of spinless fermions and speak about conductivity $\sigma$ ($\sigma$ and $D$ are trivially proportional to each other).

Because we want to study spin transport in the limit of large $\lambda$ it is crucial to have access to sufficiently large systems such that $L \gg \lambda \gg 1$. To achieve that we will use an explicit nonequilibrium driving setting where the driving is effectively accounted for by boundary Lindblad operators. The evolution of the system's density operator $\rho(t)$ is therefore described by the Lindblad master equation~\cite{Lindblad1,Lindblad2},
\begin{eqnarray}
\frac{{\rm d}\rho}{{\rm d}t}&=&{\cal L}(\rho)=\ii[\rho,H]+\sum_k 2L_k \rho L_k^\dagger-\{\rho, L_k^\dagger L_k\}.
\label{eq:Lin}
\end{eqnarray}
To force a nonzero current through the system and effectively describe driving we use 4 Lindblad operators $L_k$ acting on the boundary spins, $L_1=\sqrt{(1+\mu)}\,\sigma^+_1, L_2= \sqrt{(1-\mu)}\, \sigma^-_1$ and $L_3 =  \sqrt{(1-\mu)}\,\sigma^+_{L}, L_4= \sqrt{(1+\mu)}\, \sigma^-_{L}$. After a long time the solution of the Lindblad equation converges to a nonequilibrium steady state (NESS) $\rho_\infty$. If the driving parameter is $\mu=0$ the steady-state is a trivial $\rho_\infty \propto \mathbbm{1}$ as such driving represents an equilibrium driving at infinite temperature. For finite $\mu$ though there will be a nonzero magnetization gradient and a current in the NESS. Specifically, we are interested in the NESS expectation value of the local spin, $z_k = \tr{(\rho_\infty \sz{k})}$, and of the spin current $J=2\tr{(\rho_\infty(\sx{k}\sy{k+1}-\sy{k}\sx{k+1}))}$. Due to the continuity equation the current $J$ is independent of the site $k$. We will use $\mu=0.1$ which is in the linear regime in which $z_k$ and $J$ are proportional to $\mu$. Such driving has been used many times~\cite{dario21} in the last decade to study transport and is quite efficient, sometimes enabling numerical calculation of the NESS $\rho_\infty$ for systems with $L \sim 10^3$. For more numerical details see the Appendix.

The main object of our study is the scaling of the NESS current $J$ with system size $L$ and $\lambda$. Let us first recall the definition of the dynamical scaling exponent $z$. In a closed setting, that is without reservoirs, one has a scaling relation between distance and time as $x \sim t^{1/z}$, e.g., the variance of a localized disturbance will grow as $(\delta x)^2 \sim t^{2/z}$ with time. For instance, if one deals with diffusive transport one will have $z=2$, if one has ballistic then $z=1$. In a nonequilibrium situation where one explicitly drives the system the scaling exponent $z$ will be instead reflected in the scaling of the steady-state current $J$ with system size (see e.g. the review in Ref.~\onlinecite{dario21}). In the linear response regime, where the current is proportional to the driving potential difference $\mu$, one has  
\begin{equation}
J \propto \frac{\mu}{L^{z-1}}.
\label{eq:JL}
\end{equation}
The power of the algebraic scaling of current with $L$ therefore defines the transport type. In the case of diffusion where $z=2$ the proportionality coefficient $D$ is the diffusion constant,
\begin{equation}
J=D \frac{2\mu}{L}.
\end{equation}

\begin{figure}[t!]
  \centerline{\includegraphics[width=3.2in]{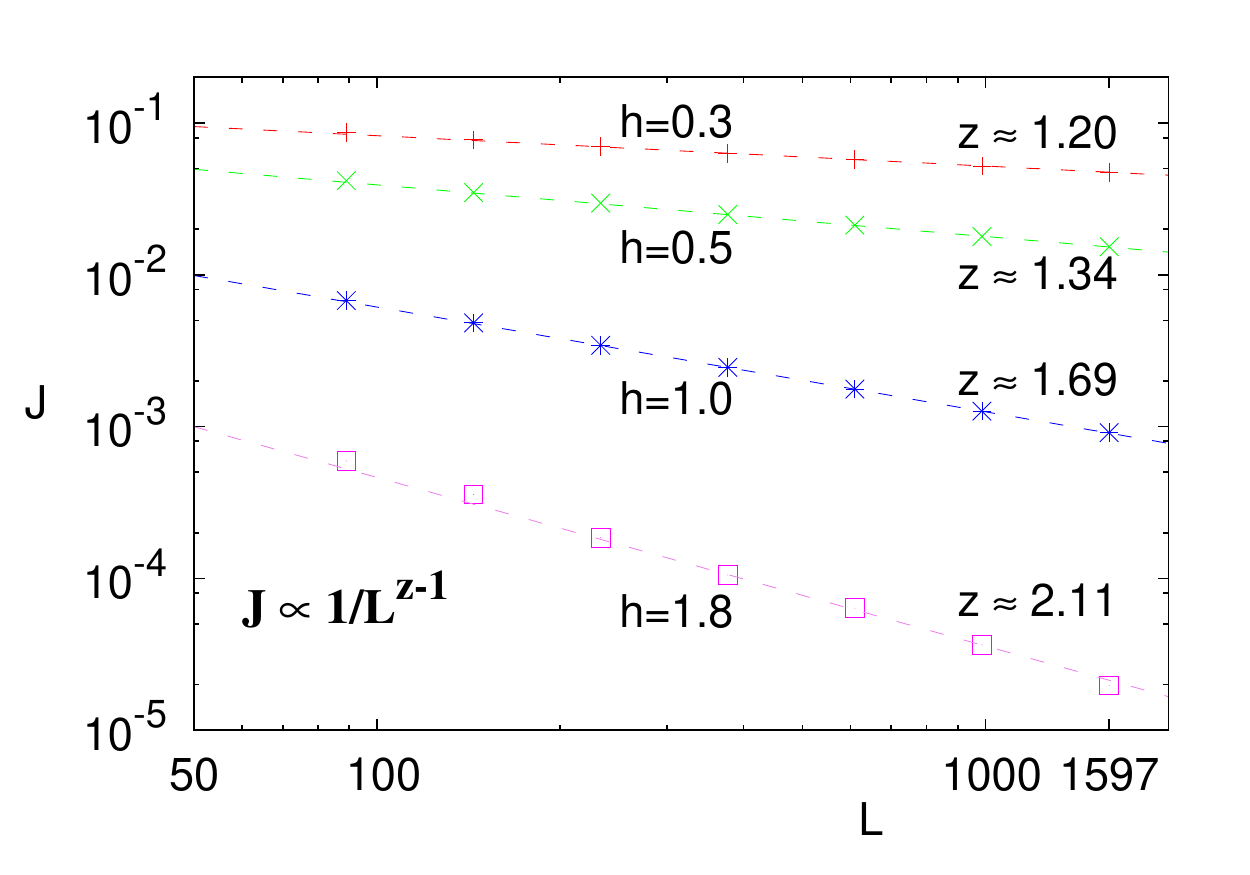}}
  \caption{Anomalous transport in the non-interacting Fibonacci model, Eq.~(\ref{eq:H}) with $V_j\equiv 0$. Dashed lines are $\sim 1/L^{z-1}$ with the best fitting powers $z$.}
\label{fig:free}
\end{figure}

To check the modified Matthiessen's rule we need the dynamical scaling exponent $z$ of the clean non-interacting model with all $V_j \equiv 0$. This has been studied many times, beginning with Ref.~\onlinecite{HiramotoAbe}; here we for completeness numerically calculate $z$ using the same Lindblad NESS driving that we then use for the interacting case. In Fig.~\ref{fig:free} we show these data. Because the model is non-interacting one can in fact avoid going through the matrix product operator (MPO) ansatz and tDMRG to obtain $\rho_\infty$, and numerically treat very large systems ($L > 10^4$) more directly~\cite{fractalNESS}. However, we show only systems upto $L=1597$ as this will be the largest size that we will be able to simulate in the interacting model.
Dynamical exponents $z$ obtained from a boundary driven Lindblad setting reported in Fig.~\ref{fig:free} are within $5\%$ of the exponents obtained from unitary dynamics~\cite{foot2} in Ref.\onlinecite{fibo19}.

\section{Modified Matthiessen's rule}

\begin{figure}[t!]
  \centerline{\includegraphics[width=3.3in]{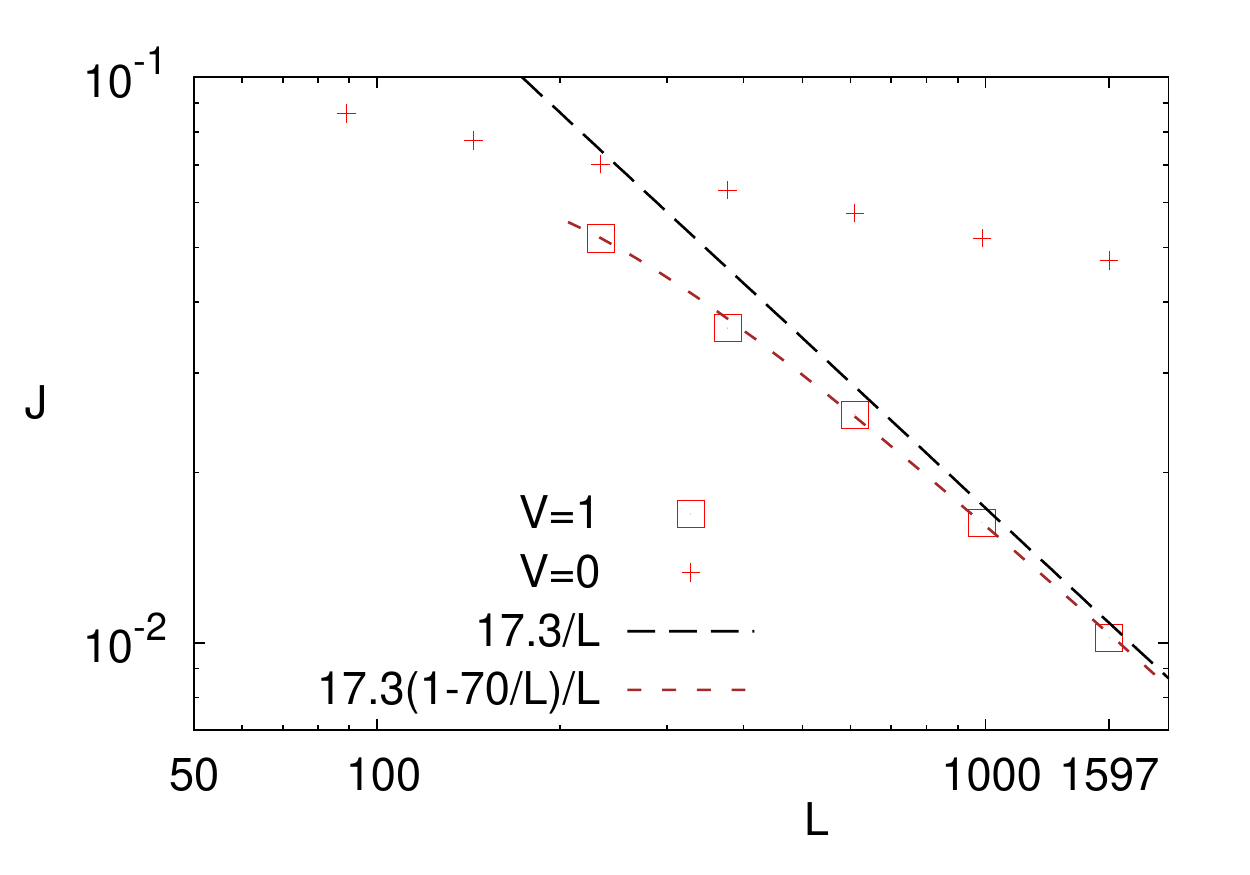}}
  \caption{Scaling of the NESS current $J$ with system size $L$ in the Fibonacci model with $\lambda=128$, and $h=0.3$. While the non-interacting model is superdiffusive with $z\approx 1.2$ (red pluses, the same data as in Fig.~\ref{fig:free}), the interacting one (squares) is diffusive with $J \propto 1/L$. Dashed black and brown curves are the best fitting leading and subleading asymptotics.}
\label{fig:jh03}
\end{figure}

We now study the interacting model for different values of $h$ and distances $\lambda$ between sites with interaction $V_j=1$. Expectedly, in all cases studied transport is diffusive in the TDL. An example of data demonstrating that is shown in Fig.~\ref{fig:jh03}.
We can see that for sufficiently large $L$ one gets diffusive scaling $J \sim 1/L$. From the brown dashed curve that overlaps with numerical points we can also see that a relative finite-size correction behaves as $\sim 1/L$ as is expected theoretically for diffusive boundary-driven systems~\cite{nessKubo}. For the shown large $\lambda=128$ the correction $\approx 70/L$ is also rather large; one needs $L>700$ in order to be within $10\%$ of the asymptotic $J \sim 17.3/L$. The diffusion constant can be read from the prefactor and is therefore equal to $D \approx 17.3/(2\mu) \approx 86$.

\begin{figure}[t!]
  \centerline{\includegraphics[width=3.2in]{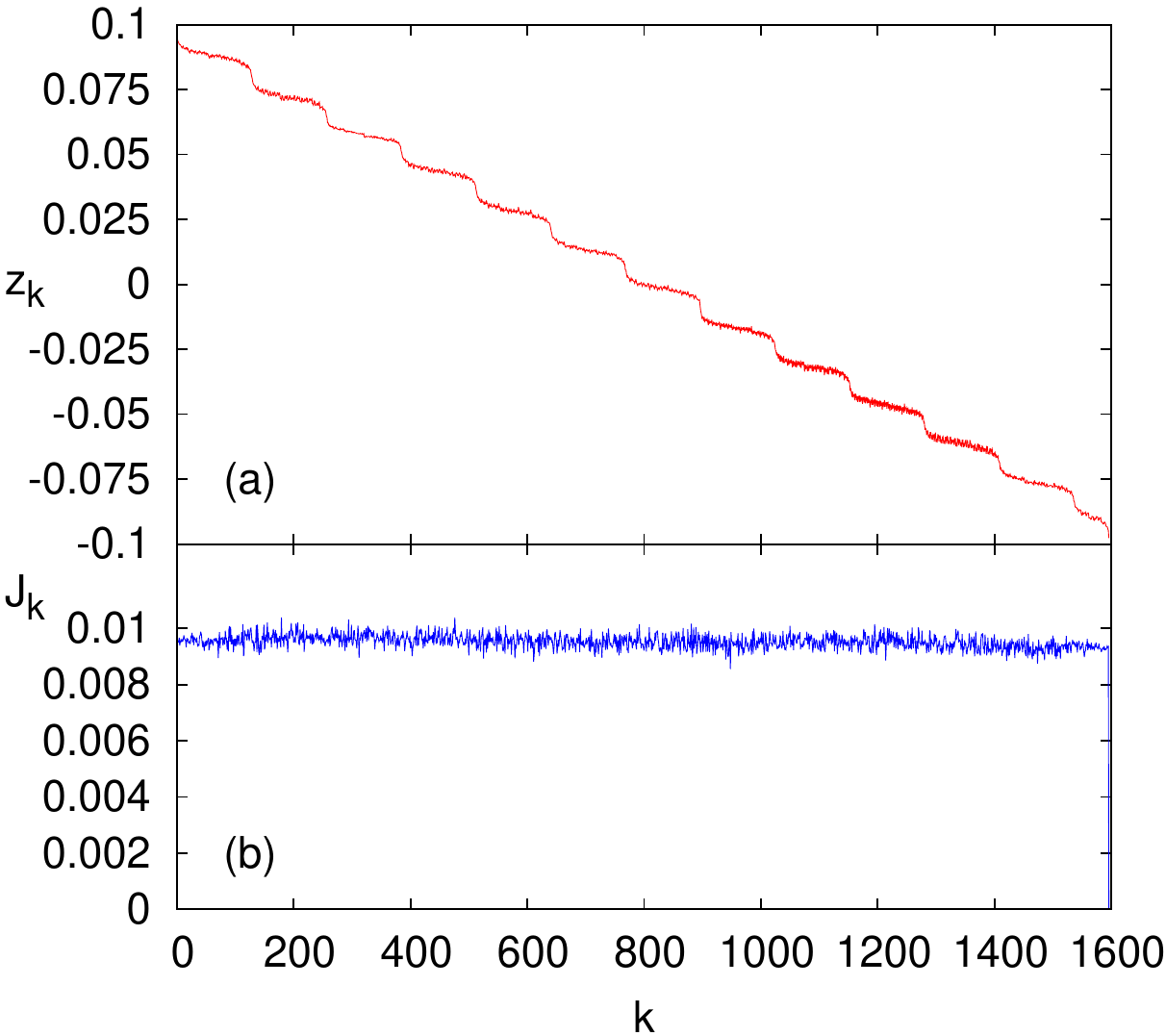}}
  \caption{NESS expectation value of local spin (a), and local spin current (b) for $L=1597$, $\lambda=128$, and $h=0.3$. In (a) we can see a repeating pattern of noninteracting sections of length $\lambda$ separated by bonds with impurities.}
\label{fig:zh03}
\end{figure}

Let us derive the theoretical prediction for the scaling of the diffusion constant (or, equivalently, of conductivity) with $\lambda$. In Fig.~\ref{fig:zh03} we show the NESS magnetization profile and the current for one set of parameters. 
The NESS current is apart from noisy fluctuations due to finite matrix product operator (MPO) size $\chi$ homogeneous. The size of the fluctuations is one way of estimating the error of $J$ (about $5 \%$ in this case, see Table~\ref{tab:numerics}). More telling is the spin profile. We can see that in-between the sites with scattering the profile is reminiscent of a one in an anomalous model. The jump in the magnetization (i.e., in the driving potential) across a segment with resistance $R$ is $J\cdot R$. The total jump in magnetization across the whole chain, which is $2\mu$ in our case, is simply a sum of jumps at the impurity sites and jumps across the non-interacting segments. For a localized perturbation, in our case a single bond, we can assume that it has a finite resistance $R_0$. The jump in magnetization at each site with perturbation will therefore scale as $\sim J R_0$ and will in the TDL go to zero for all systems with $z>1$ because $J \sim 1/L^{z-1}$. Therefore, in the TDL all magnetization drop occurs in the scattering-free non-interacting segments of length $\lambda$. There are $K=L/\lambda$ such segments so that in each magnetization will change by $\Delta z=2\mu/K$. Looking at a non-interacting segment of length $\lambda$ that is described by the scaling exponent $z$, we can conclude that the current should be $J=\Delta z/\lambda^{z-1}$. This brings us to the final result~\cite{prl20} that in the TDL one has $J \sim 2\mu \lambda/(L \lambda^{z-1})$, giving diffusion constant scaling for large $\lambda$
\begin{equation}
D \propto \frac{1}{\lambda^{z-2}}.
\label{eq:mod}
\end{equation}  
This is the modified Matthiessen's rule. Only for the ballistic clean model does one recover the standard scaling $D \sim \lambda \sim 1/K$ from Eq.~(\ref{eq:lin}). Because the derivation is completely general, the only ingredient really being a defining relation of anomalous transport in a clean model, Eq.~(\ref{eq:JL}), it is expected to hold in the limit $\lambda\to \infty$ for any system that has anomalous transport and to which one adds dilute impurities at average linear distance $\lambda$.

A relevant question is how common are such anomalous systems in which the rule has to be modified? One might argue that it is rather special; typically one expects to have either a ballistic transport ($z=1$) in integrable models~\cite{zotos97}, like in e.g. homogeneous free fermions, or a diffusive one ($z=2$) in the case of generic interacting systems. While the Fibonacci potential is clearly special one can note that in non-interacting models one can in general engineer the system's properties~\cite{mantica97}. Recently it has become clear that anomalous transport though is not limited only to non-interacting quantum systems. Specifically, superdiffusive $z=3/2$ behavior can be found in interacting systems like the isotropic Heisenberg model~\cite{znidaric11} of real-material relevance~\cite{Kawamata08,Hlubek11,Hlubek12}, as well as in other integrable isotropic~\cite{foot1} systems~\cite{dupont20,vir20b,sid20,enej20,rahul21,enej21}, including classical ones~\cite{bojan13,ziga19,manas19}. Another case, not completely resolved, is a possible more generic superdiffusion emerging from effective theories at low temperature~\cite{vir20,marko20,rahul21}. Anomalous transport can also be engineered in stochastic models~\cite{slava}.

What is the crux that leads to the modified rule in which the resistivity is not simply proportional to the number of impurities? It is the clean sections without impurities that for large $\lambda$ and in the TDL can not be neglected. While it is the extensive number of scattering sites that causes diffusive scaling of current $J \sim 1/L$, it is the anomalous parts in-between that determine the value of $D$. If one would just add resistances of all $K$ scattering sites and $K$ anomalous segments one would get a total chain resistance $KR_0+K \lambda^{z-1}=L \lambda^{z-2}(R_0/\lambda^{z-1}+1)$, so that for $z>1$ and large $\lambda$ one can neglect the scattering term with $R_0$. Because the clean sections are sub-ballistic they are not negligible, in fact, they dominate over scattering on impurities. One can say that in a way additivity still holds if one understands it correctly: it should be applied to the anomalous parts instead of only summing up the scattering rates on impurities. We can also see that having large $\lambda$ while keeping the scattering sections at a fixed length (fixed $R_0$) is crucial. If we would also increase the length of the scattering sections with $L$ one would recover the original Matthiessen's rule where $D \sim \lambda$.

\begin{figure}[t!]
  \centerline{\includegraphics[width=3.3in]{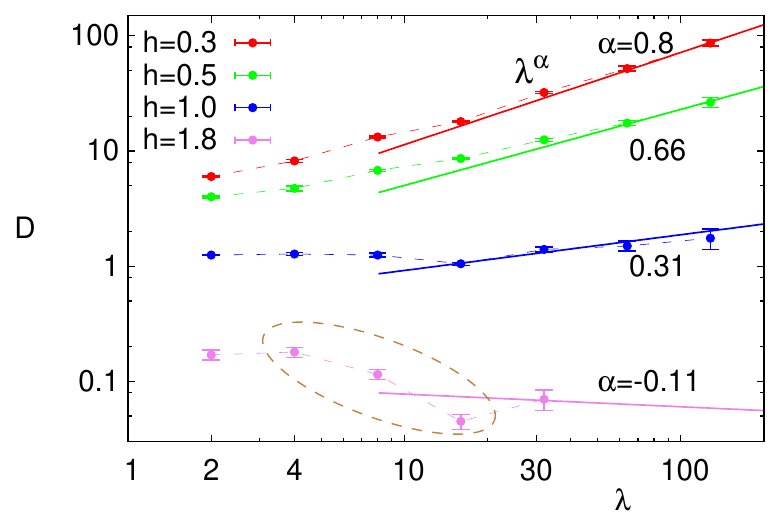}}
  \caption{Dependence of the spin diffusion constant $D$ on the distance $\lambda$ between impurities. Straight lines are $\sim \lambda^\alpha$, with powers $\alpha$ being equal to theoretical $\alpha=2-z$, with $z$ from Fig.~\ref{fig:free}. The modified Matthiessen's rule (\ref{eq:mod}) holds for large $\lambda>30$, whereas for small $\lambda$ and large $h=1.8$ (circled points), where the rule is not expected to hold, one has a regime where more impurities produces larger $D$ (``more is less'' -- more scattering at smaller $\lambda$ causes less resistance).}
\label{fig:D}
\end{figure}

Let us now verify if the modified Matthiessen's rule (\ref{eq:mod}) indeed holds for any value of $z$, not just $z=3/2$ that has been already checked in Ref.~\onlinecite{prl20}. Doing plots like in Fig.~\ref{fig:jh03} we calculate $D$ for a range of $\lambda$ and 4 different values of the on-site field amplitude $h$ thereby tuning the value of $z$ in the non-interacting clean model without impurities. Results are shown in Fig.~\ref{fig:D}.
We can see that for all superdiffusive cases, $h=0.3, 0.5, 1.0$, and for large enough $\lambda$ (about $\lambda >30$) we indeed get the modified scaling $D \sim \lambda^{2-z}$ with theoretical $z$ from Fig.~\ref{fig:free}. Unfortunately, for subdiffusive $h=1.8$, where one would expect that $D$ would decrease with $\lambda$ at a sufficiently large $\lambda$, the numerics gets very hard and we could not get sufficiently precise results at larger $\lambda$ (errors are large and we only have data for $\lambda \le 32$ which is likely not yet in the true asymptotic regime of large $\lambda$; see the Appendix~\ref{app:A}). Nevertheless, data for $h=1.8$ in Fig.~\ref{fig:D} are compatible with the expected asymptotic $D \sim 1/\lambda^{0.1}$. 

There is, however, one other interesting behavior visible at smaller $\lambda$, where the modified Matthiessen's rule (\ref{eq:mod}) does not yet hold. For $h\le 1.0$ we can see (Fig.~\ref{fig:D}) that for such small $\lambda$ (say $\lambda=4-10$) one has the expected behavior: putting more impurities in our system, i.e., decreasing $\lambda$, the diffusion constant decreases. Heuristically one can explain this decrease in diffusion as being due to an increased scattering. However for $h=1.8$ and $\lambda \approx 4-16$ the diffusion constant instead increases as one decreases $\lambda$ (a small effect of a similar kind is visible also for $h=1.0$). For instance, at $\lambda=16$ one has diffusion constant $D \approx 0.05$; adding then twice as many impurities, resulting in $\lambda=8$, one would expect that $D$ would decrease, but instead it increases to $D \approx 0.12$. In this regime more scattering results in less resistivity. One could argue that this is indeed in-line with Eq.(\ref{eq:mod}) for subdiffusive clean systems with $z>2$, however we stress that the rule is not yet expected to hold at such small $\lambda$. Also, the decrease in $D$ is much larger than would be predicted by Eq.(\ref{eq:mod}). It is true though that the origin of this effect could be similar as for the modified Matthiessen's rule but with a more complicated dependence on $\lambda$ due to not-yet asymptotic $\lambda$. We also note that the effect is similar in spirit to various noise (dephasing) assisted enhancements of transport observed in e.g. Refs.~\onlinecite{plenio08,horvat13,lacerda21,cecilia21}.

\section{Conclusion}

We studied how transport coefficients like conductivity or the diffusion constant depend on the concentration of localized integrability-breaking perturbations -- impurities. The standard textbook argument would suggest that in the limit of dilute impurities the diffusion constant will be inversely proportional to the number of impurities, also known as Matthiessen's rule. We have demonstrated that this rule is in fact true only in a special case when a system without impurities is a ballistic conductor. In the generic situation of anomalous transport the scaling has to be instead modified, so that the diffusion constant has a non-trivial power-law dependence on the impurity density, with the power being given by the dynamical exponent. This modified rule has been verified for spin transport at infinite temperature in the interacting Fibonacci model. We also found an interesting non-asymptotic regime in which adding more impurities can increase the diffusion constant.

What we were not able to fully check is the case of subdiffusion. Our numerical results are for a specific type of impurities -- interaction on every $\lambda$-th bond -- and for a particular conserved charge (spin). We expect the modified Matthiessen's rule to hold in general, also if one would start with a clean anomalous nonintegrable system.

\section*{Acknowledgments}

Support from Grants No.~J1-1698 and No.~P1-0402 from the Slovenian Research Agency is acknowledged.

\appendix
\section{Numerical details}
\label{app:A}

The numerical method used to obtain the NESS $\rho_\infty$ of the Lindblad equation is the same one as used in our previous works. Here we briefly repeat the essentials (for more see e.g. the references cited in Ref.~\onlinecite{znidaric11,Max20,diff}) and give the representative parameters used.

The density operator $\rho(t)$ is expanded in the basis of products of Pauli matrices $\sigma^{\pmb{\alpha}}=\sigma_1^{\alpha_1} \cdots \sigma_L^{\alpha_L}$ as,
\begin{equation}
\rho(t)=\sum_{\pmb{\alpha}} c_{\pmb{\alpha}}(t) \sigma^{\pmb{\alpha}},
\end{equation}
where $\sigma_k^{\alpha_k} \in \{ \sx{k},\sy{k},\sz{k},\mathbbm{1}_k\}$. Expansion coefficients are written in terms of the product of $\chi \times \chi$ dimensional matrices $M$ (MPO),
\begin{equation}
c_{\pmb{\alpha}}= \langle M_1^{(\alpha_1)} \cdots M_L^{(\alpha_L)} \rangle.
\end{equation}
The time-dependent solution $\rho(t)$ is then obtained as $\rho(t)={\rm e}^{{\cal L}t} \rho(0)$ by evolving matrices $M$ in time using small Trotter-Suzuki timesteps of length $\Delta t=0.05$, using standard procedures as in e.g. pure-state tDMRG~\cite{Schollwock}. Note that because the local operator basis is of size $4$, as opposed to $2$ for a pure-state evolution of qubit chains, the complexity of simulating $\rho(t)$ in an $L$-site qubit chain is the same as simulating pure states in a ladder of length $L$ (or a spin-$3/2$ chain of length $L$).

%\begingroup
%\squeezetable
\begin{table}[tp]
\begin{ruledtabular}
\begin{tabular}{rccccc} 
 $h$ & $\lambda$ & $L$ & $\chi$ & $t_\infty$ & err.$J$ [\%] \\
\midrule
%\multicolumn{1}{l}{0.3}\\
0.3 & 8 & 233 & 100 & $7\cdot 10^2$ &  1 \\
 &   & 610 & 100 & $2\cdot 10^3$ & 3 \\
 & 128 & 610 & 100 & 5 $\cdot 10^3$ & 5 \\
 &   & 1597 & 100 & $4\cdot 10^3$ & 5 \\
%\multicolumn{1}{l}{0.5}\\
0.5 & 32 & 233 & 80 & $3\cdot 10^2$ & 2\\
&    & 610 & 50 &$7\cdot 10^2$ & 7\\
& 128 & 610 & 200 &$8\cdot 10^2$ & 4\\
&    & 1597 & 100 &$2\cdot 10^3$ & 13\\
%\multicolumn{1}{l}{1.0}\\
1.0 & 8 & 233 & 150 &$8\cdot 10^2$ & 3\\
&    & 610 & 200 &$2\cdot 10^3$ & 2\\
& 32 & 144 & 100 &$1\cdot 10^3$ & 6\\
&    & 233 & 200 &$2\cdot 10^3$ & 10\\
%\multicolumn{1}{l}{1.8}\\
1.8 & 8 & 233 & 100 &$1\cdot 10^4$ & 10\\
&    & 610 & 100 &$5\cdot 10^3$ & 7\\
& 32 & 233 & 200 &$5\cdot 10^3$ & 10\\
&    & 987 & 100 &$1\cdot 10^4$ & 20
\end{tabular}
\end{ruledtabular}
\caption{MPO size $\chi$, convergence times $t_\infty$ to the NESS, and the estimated error of the steady-state current $J$ for a couple of system sizes $L$, distances between impurities $\lambda$ and on-site fields $h$. Not all $L$ used in Fig.~\ref{fig:D} are shown.}
\label{tab:numerics}
\end{table}
%\endgroup
The NESS is in our case always unique so one can start with an arbitrary initial $\rho(0)$, eventually converging to the NESS after a long time. The crucial parameter is the matrix size $\chi$. Namely, the larger the $\chi$ the smaller are truncation errors, however, the complexity of each step grows as $\sim \chi^3$ and therefore simulations get very slow at larger $\chi$. Even if we start with a product initial density operator which requires only $\chi=1$ evolution will cause the necessary $\chi$ to quickly grow with time. Therefore we simply keep the matrix size constant and equal to $\chi$ from the very beginning. Because the Lindbladian propagator is not unitary, the orthogonality of the Schmidt eigenvectors is not preserved. To remedy that we re-orthogonalize our MPO representation of $\rho(t)$ every few steps (typically 10-20 steps, when we also calculate the expectation values of the magnetization and the local current).

The required convergence time $t_\infty$ until the NESS is reached as well as the necessary $\chi$ for a given precision greatly varies with the potential strength $h$ and with system size $L$. Some representative numbers can be found in Table~\ref{tab:numerics}. In particular, and inline with previous studies of the Heisenberg model with random fields~\cite{Max20}, computational complexity rapidly increases with growing $h$. For $h$ large enough so that the clean non-interacting system is subdiffusive (e.g., $h=1.8$ where $z\approx 2.11 > 2$), the convergence time as well as the estimated errors at fixed $\chi$, say $\chi=100$, rapidly increases. This means that we unfortunately could not reliably probe the case of large $z$ and therefore a regime of negative powers $2-z$. Solving a boundary driven Lindblad equation with an MPO ansatz is therefore a very good method at small $h$, but becomes less efficient at larger on-site fields~\cite{Max20}.

% FIBO PRB'19 Fig.2 beta=1/z= 0.87(z=1.15)@0.3, 0.78(z=1.28)@0.5, 0.61(z=1.64)@1, 0.48(z=2.08)@1.8
% enDiagFib gamma=z-1= 0.20@0.3, 0.34@0.5, 0.69@1, 1.11@1.8


\begin{thebibliography}{}

\bibitem{Kittel} C.~Kittel, {\em Introduction to solid state physics}, p.161 (John Wiley \& Sons, 1996).
% at low T by impurities. rates are additive, i.e., resistivities are additive, e.g. total resistance is a sum due to phonons and impurities.

\bibitem{Mermin} N.~W.~Ashcrotf and N.~D.~Mermin, {\em Solid state physics}, p.323 (Harcourt College Publishing, 1976).
% R_12>R_1 + R_2

\bibitem{matthiessen} A.~Matthiessen and C.~Vogt, \tit{Ueber den Einflus der Temperatur auf die elektrische Leitungsf\" ahigkeit der Legirungen} Ann.~Phys. {\bf 198}, 19 (1884)


\bibitem{Ziman} J.~M.~Ziman, {\em Electrons and phonons}, p.285 (Oxford University Press, 1960).
% two independent mechanisms -> probabilities should add (Fermi's Golden Rule) -> rates additive -> resistance additive

\bibitem{Bass} J.~Bass, \tit{Deviations from Matthiessen's rule} Advances in Physics {\bf 21}, 431 (1972).
% an early review; (skoraj?)vsi podatki od T, ne od koncentracije


\bibitem{Klein01} L.~Klein, \etal{Y.~Kats, N.~Wiser, M.~Konczykowski, J.~W.~Reiner, T.~H.~Geballe, M.~R.~Beasley, and A.~Kapitulnik}, \tit{Negative deviations from Matthiessen's rule for SrRuO$_3$ and CaRuO$_3$} Europhys.~Lett. {\bf 55}, 532 (2001).
% irradiation SrRuO_3, Fig.2: naklon premice rho(c) se malo manjsa z dozo c (za 20%).

\bibitem{prl20} M.~\v Znidari\v c, \tit{Weak integrability breaking: chaos with integrability signature in coherent diffusion} Phys.~Rev.~Lett. {\bf 125}, 180605 (2020).


\bibitem{Kawamata08} T.~Kawamata \etal{, N.~Takahashi, T.~Adachi, T.~Noji, K.~Kudo, N.~Kobayashi, and Y.~Koike}, \tit{Evidence for ballistic thermal conduction in the one-dimensional $S=1/2$ Heisenberg antiferromagnetic spin system ${\rm Sr}_2{\rm CuO}_3$} J.~Phys.~Soc.~Jpn. {\bf 77}, 034607 (2008).
% Pd doped (prekine verigo), two x=0,0.004,0.01; mag.suscept in heat cond., oba konsistentno L_spin 470->800 oz. 1/x^0.58

\bibitem{Hlubek11} N.~Hlubek \etal{R.~Saint-Martin, S.~Nishimoto, A.~Revcolevschi, S.-L.~Drechsler, G.~Behr, J.~Trinckauf, J.~E.~Hamann-Borrero, J.~Geck, B.~B\" uchner, and C.~Hess}, \tit{Bond disorder and breakdown of ballistic heat transport in the spin-1/2 antiferromagnetic Heisenberg chain as seen in Ca-doped SrCuO$_2$} Phys.~Rev.~ B {\bf 84}, 214419 (2011); A.~Mohan \etal{N.~Sekhar Beesetty, N.~Hlubek, R.~Saint-Martin, A.~Revcolevschi, B.~B\" uchner, and C.~Hess}, \tit{Bond disorder and spinon heat transport in the $S = \frac{1}{2}$ Heisenberg spin chain compound Sr$_2$CuO$_3$: From clean to dirty limits} Phys.~Rev.~ B {\bf 89}, 104302 (2014).
% Ca doped, ki zamenja Cu (=bond disorder/off-diag) SrCuO2 AFM XXX: se ujema z 1/x
% PRB'14 single-chain Sr_2CuO_3 in podobni zakljucki

\bibitem{Hlubek12} N.~Hlubek, X.~Zotos, S.~Singh, R.~Saint-Martin, A.~Revcolevschi, B.~B\" uchner, and C.~Hess, \tit{Spinon heat transport and spin-phonon interaction in the spin-1/2 Heisenberg chain cuprates Sr$_2$CuO$_3$ and SrCuO$_2$} J.~Stat.~Mech {\bf 2012}, P03006 (2012).
% 2N,3N,4N ; thermal conductivity; za 4N imajo l_mag=0.54\mu m(=1300 mreznih), za prejsnje podatke pa 3N=0.2, 2N(99%pure)=0.05


\bibitem{Lea} L.~F.~Santos, \tit{Integrability of a disordered Heisenberg spin-1/2 chain} J.~Phys.~A {\bf 37}, 4723 (2004).

\bibitem{Brenes} M.~Brenes, E.~Mascarenhas, M.~Rigol, and J.~Goold, \tit{High-temperature transport in the XXZ chain in the presence of an impurity} Phys.~Rev.~B {\bf 98}, 235128 (2018).

\bibitem{speck} L.~F.~Santos, F.~Perez-Bernal, and E.~J.~Torres-Herrera, \tit{Speck of chaos} Phys.~Rev.~Res. {\bf 2}, 043034 (2020).


\bibitem{rosch00} A.~Rosch and N.~Andrei, \tit{Conductivity of a clean one-dimensional wire} Phys.~Rev.~Lett. {\bf 85}, 1092 (2000).

\bibitem{jung06} P.~Jung, R.~W.~Helmes, and A.~Rosch, \tit{Transport in almost integrable models: perturbed Heisenberg chains} Phys.~Rev.~Lett. {\bf 96}, 067202 (2006).

\bibitem{huang13} Y.~Huang, C.~Karrasch, and J.~E.~Moore, \tit{Scaling of electrical and thermal conductivities in an almost integrable chain} Phys.~Rev.~B {\bf 88}, 115126 (2013).

\bibitem{robin14} R.~Steinigeweg, F.~Heidrich-Meisner, J.~Gemmer, K.~Michielsen, and H.~De Raedt, \tit{Scaling of diffusion constants in the spin-1/2 XX ladder} Phys.~Rev.~B {\bf 90}, 094417 (2014).

\bibitem{vadim18} R.~J.~Sanchez, V.~K.~Varma, and V.~Oganesyan, \tit{Anomalous and regular transport in spin-1/2 chains: ac conductivity} Phys.~Rev.~B {\bf 98}, 054415 (2018).

\bibitem{pnas18} M.~\v Znidari\v c and M.~Ljubotina, \tit{Interaction instability of localization in quasiperiodic systems} Proc.~Natl.~Acad.~Sci.~USA {\bf 115}, 4595 (2018).

\bibitem{comment21} M.~\v Znidari\v c, \tit{Comment on ``Nonequilibrium steady state phases of the interacting Aubry-Andr\' e-Harper model''} Phys.~Rev.~B {\bf 103}, 237101 (2021).


\bibitem{diff} J.~S.~Ferreira and M.~Filippone, \tit{Ballistic-to-diffusive transition in spin chains with broken integrability} Phys.~Rev.~B {\bf 102}, 184304 (2020).

\bibitem{Friedman20} A.~J.~Friedman, S.~Gopalakrishnan, and R.~Vasseur, \tit{Diffusive hydrodynamics from integrability breaking} Phys.~Rev.~B {\bf 101}, 180302(R) (2020).
% hydro, dobis difuzijo

  
\bibitem{Doyon} J.~Durnin, M.~J.~Bhaseen, and B.~Doyon, \tit{Non-equilibrium dynamics and weakly broken integrability} Phys.~Rev.~Lett. {\bf 127}, 130601 (2021).
%arXiv:2004.11030 (2020).
% 
  
\bibitem{Vasseur} J.~Lopez-Piqueres, B.~Ware, S.~Gopalakrishnan, and R.~Vasseur, \tit{Hydrodynamics of nonintegrable systems from a relaxation-time approximation} Phys.~Rev.~B {\bf 103}, 060302 (2020). 
% relax.time approx. 

\bibitem{kohmoto83} M.~Kohmoto, L.~P.~Kadanoff, and C.~Tang, \tit{Localization problem in one dimension: mapping and escape} Phys.~Rev.~Lett. {\bf 50}, 1870 (1983).

\bibitem{ostlund83} S.~Ostlund, R.~Pandit, D.~Rand, H.~J.~Schellnhuber, and E.~D.~Siggia, \tit{One-dimensional Schrödinger equation with an almost periodic potential} Phys.~Rev.~Lett. {\bf 50}, 1873 (1983).

\bibitem{kitaev86} P.~A.~Kalugin, A.~Y.~Kitaev, and L.~S.~Levitov, \tit{Electron spectrum of a one-dimensional quasicrystal} Sov.~Phys.~JETP {\bf 64}, 410 (1986).

\bibitem{sutherland87} B.~Sutherland and M.~Kohmoto, \tit{Resistance of a one-dimensional quasicrystal: power-law growth} Phys.~Rev.~B {\bf 36}, 5877 (1987).

\bibitem{HiramotoAbe} H.~Hiramoto and S.~Abe, \tit{Dynamics of an electron in quasiperiodic systems. I. Fibonacci model} J.~Phys.~Soc.~Japan {\bf 57}, 230 (1988).  

\bibitem{damanik} D.~Damanik, M.~Embree, and A.~Gorodetski, \tit{Spectral properties of Schr\" odinger operators arising in the study of quasicrystals}, In: J.~Kellendonk, D.~Lenz, J.~Savinien (eds) {\em Mathematics of Aperiodic Order}. Progress in Mathematics, {\bf 309} p.307-370 (Birkh\" auser, 2015).
%Basel, https://doi.org/10.1007/978-3-0348-0903-0_9


\bibitem{vidal} J.~Vidal, D.~Mouhanna, and T.~Giamarchi, \tit{Correlated fermions in a one-dimensional quasiperiodic potential} Phys.~Rev.~Lett. {\bf 83}, 3908 (1999); J.~Vidal, D.~Mouhanna, and T.~Giamarchi, \tit{Interacting fermions in self-similar potentials} Phys.~Rev.~B {\bf 65}, 014201 (2001).
% low-T RG

\bibitem{lacerda21} A.~M.~Lacerda, J.~Goold, and G.~T.~Landi, \tit{Dephasing enhanced transport in boundary-driven quasiperiodic chains} Phys.~Rev.~B {\bf 104}, 174203 (2021).

\bibitem{akkermans14} D.~Tabese \etal{, E.~Gurevich, F.~Baboux, T.~Jacqmin, A.~Lemaitre, E.~Galopin, I.~Sagnes, A.~Amo, J.~Bloch, and E.~Akkermans}, \tit{Fractal energy spectrum of a polariton gas in a Fibonacci quasiperiodic potential} Phys.~Rev.~Lett. {\bf 112}, 146404 (2014).

\bibitem{strkalj20} V.~Golobot \etal{, A.~\v Strkalj, N.~Pernet, J.~L.~Lado, C.~Dorow, A.~Lemaitre, L.~Le Gratiet, A.~Harouri, I.~Sagnes, S.~Ravets, A.~Amo, J.~Bloch, and O.~Zilberberg}, \tit{Emergence of criticality through a cascade of delocalization transitions in quasiperiodic chains} Nat.~Phys. {\bf 16}, 832 (2020).
% interpolirajo AAH-Fibo, cavity-polariton exper

\bibitem{Alet18} N.~Mace, N.~Laflorencia, and F.~Alet, \tit{Many-body localization in a quasiperiodic Fibonacci chain} SciPost Phys. {\bf 6}, 050 (2019).

\bibitem{fibo19} V.~K.~Varma and M.~\v Znidari\v c, \tit{Diffusive transport in a quasiperiodic Fibonacci chain: Absence of many-body localization at weak interactions} Phys.~Rev.~B {\bf 100}, 085105 (2019).

\bibitem{goold21} C.~Chiaracane, F.~Pietracaprina, A.~Purkayastha, and J.~Goold, \tit{Quantum dynamics in the interacting Fibonacci chain} Phys.~Rev.~B {\bf 103}, 184205 (2021).
% interacting, typicality (L<=24), wavepacket spreading, za vecje V subdiff.

\bibitem{doggen21} A.~\v Strkalj, E.~V.~H.~Doggen, I.~V.~Gornyi, and O.~Zilberberg, \tit{Many-body localization in the interpolating Aubry-Andr{\' e}-Fibonacci model} Phys.~Rev.~Research {\bf 3}, 033257 (2021).
% arXiv:2106.13841 (2021).

\bibitem{Lindblad1} V.~Gorini, A.~Kossakowski, and E.~C.~G. Sudarshan, \tit{Completely positive dynamical semigroups of N-level systems} J.~Math.~Phys. {\bf 17}, 821 (1976).

\bibitem{Lindblad2} G.~Lindblad, \tit{On the generators of quantum dynamical semigroups} Commun.~Math.~Phys. {\bf 48}, 119 (1976).

\bibitem{dario21} G.~T.~Landi, D.~Poletti, and G.~Schaller, \tit{Non-equilibrium boundary driven quantum systems: models, methods and properties} arXiv:2104.14350 (2021).

\bibitem{fractalNESS} V. K. Varma, C. de Mulatier, and M. {\v Znidari\v c}, \tit{Fractality in nonequilibrium steady states of quasiperiodic systems} Phys. Rev. E \textbf{96}, 032130 (2017).

\bibitem{foot2} The exponents from Fig.2 of Ref.~\onlinecite{fibo19} for the domain wall as well as wavepacket spreading are $z\approx 1.15, 1.28, 1.64, 2.08$ at $h=0.3, 0.5, 1.0, 1.8$, respectively.

\bibitem{nessKubo} M.~\v Znidari\v c, \tit{Nonequilibrium steady-state Kubo formula: Equality of transport coefficients} Phys.~Rev.~B {\bf 99}, 035143 (2019).


\bibitem{zotos97} X.~Zotos, F.~Naef, and P.~Prelov\v sek, \tit{Transport and conservation laws} Phys.~Rev.~B {\bf 55}, 11029 (1997).

\bibitem{mantica97} G.~Mantica, \tit{Quantum intermittency in almost-periodic lattice systems derived from their spectral properties} Physica D {\bf 103}, 576 (1997).

\bibitem{znidaric11} M.~\v Znidari\v c, \tit{Spin transport in a one-dimensional anisotropic Heisenberg model} Phys.~Rev.~Lett. {\bf 106}, 220601 (2011).

\bibitem{foot1} Or more generally in systems with a continuous non-Abelian symmetry.

\bibitem{dupont20} M.~Dupont and J.~E.~Moore, \tit{Universal spin dynamics in infinite-temperature one-dimensional quantum magnets} Phys.~Rev.~B {\bf 101}, 121106(R) (2020).
% superdif v integ. with symm; iso S=1,3/2,2 ni ker niso int. je pa bil.-biq.

\bibitem{vir20b} V.~B.~Bulchandani, \tit{Kardar-Parisi-Zhang universality from soft gauge modes} Phys.~Rev.~B {\bf 101}, 041411(R) (2020).
% isotropic integrable magnets classical and quantum imajo 3/2


\bibitem{enej20} J.~De Nardis, S.~Gopalakrishnan, E.~Ilievski, and R.~Vasseur, \tit{Superdiffusion from emergent classical solitons in quantum spin chains} Phys.~Rev.~Lett. {\bf 125}, 070601 (2020).
% giant quasi<---classical LL solitons


\bibitem{enej21} E.~Ilievski, J.~De Nardis, S.~Gopalakrishnan, R.~Vasseur, and B.~Ware, \tit{Superuniversality of superdiffusion} Phys.~Rev.~X {\bf 11}, 031023 (2021).
% int. quantum with cont non-Ab.=3/2


\bibitem{sid20} M.~Fava \etal{, B.~Ware, S.~Gopalakrishnan, R.~Vasseur, and S.~A.~Parameswaran}, \tit{Spin crossovers and superdiffusion in the one-dimensional Hubbard model} Phys.~Rev.~B {\bf 102}, 115121 (2020).


\bibitem{rahul21} P.~Glorioso \etal{, L.~V.~Delacr\' etaz, X.~Chen, R.~M.~Nandkishore, and A.~Lucas}, \tit{Hydrodynamics in lattice models with continuous non-Abelian symmetries} SciPost Phys. {\bf 10}, 15 (2021).
% classical SU(2) chains, int. and non., are 


\bibitem{bojan13} T.~Prosen and B.~\v Zunkovi\v c, \tit{Macroscopic diffusive transport in a microscopically integrable hamiltonian system} Phys.~Rev.~Lett. {\bf 111}, 040602 (2013).

\bibitem{ziga19} \v Z.~Krajnik and T.~Prosen, \tit{Kardar-Parisi-Zhang physics in integrable rotationally symmetric dynamics on discrete space–time lattice} J.~Stat.~Phys. {\bf 179}, 110 (2020).

\bibitem{manas19} A.~Das, M.~Kulkarni, H.~Spohn, and A.~Dhar, \tit{Kardar-Parisi-Zhang scaling for an integrable lattice Landau-Lifshitz spin chain} Phys.~Rev.~E {\bf 100}, 042116 (2019).
% klasicni 3/2


\bibitem{vir20} V.~B.~Bulchandani, C.~Karrasch, and J.~E.~Moore, \tit{Superdiffusive transport of energy in one-dimensional metals} Proc.~Natl.~Acad.~Sci.~USA {\bf 117}, 12713 (2020).
% breaking of Luttinger at low-T for energy

\bibitem{marko20} J.~De Nardis, M.~Medenjak, C.~Karrasch, and E.~Ilievski, \tit{Universality classes of spin transport in one-dimensional isotropic magnets: the onset of logarithmic anomalies} Phys.~Rev.~Lett. {\bf 124}, 210605 (2020).
% tudi neint. rot. invar. bi naj imele super at low-T (likely just finite time?)
% classical Heis log^1.33(t)


\bibitem{slava} V.~Popkov, A.~Schadschneider, J.~Schmidt, and G.~M.~Sch\" utz, \tit{Fibonacci family of dynamical universality classes} Proc.~Natl.~Acad.~Sci.~USA {\bf 112}, 12645 (2015). 


\bibitem{Max20} M. \v Znidari\v c, A. Scardicchio, and V. K. Varma, \tit{Diffusive and subdiffusive spin transport in the ergodic phase of a many-body localizable system} Phys. Rev. Lett. \textbf{117}, 040601 (2016); M.~Schulz, S.~R.~Taylor, A.~Scardicchio, and M.~\v Znidari\v c, \tit{Phenomenology of anomalous transport in disordered one-dimensional systems} J.~Stat.~Mech. {\bf 2020}, 023107 (2020). 

\bibitem{Schollwock} U.~Schollw\" ock, \tit{The density-matrix renormalization group in the age of matrix product states} Annals of Physics {\bf 326}, 96 (2011).

\bibitem{plenio08} M.~B.~Plenio and S.~F.~Huelga, \tit{Dephasing-assisted transport: quantum networks and biomolecules} New J.~Phys. {\bf 10}, 113019 (2008).

\bibitem{horvat13} M.~\v Znidari\v c and M.~Horvat, \tit{Transport in a disordered tight-binding chain with dephasing} Eur.~Phys.~J.~B {\bf 86}, 67 (2013).

\bibitem{cecilia21} C.~Chiaracane, A.~Purkayastha, M.~T.~Mitchison, and J.~Goold, \tit{Dephasing-enhanced performance in quasiperiodic thermal machines} arXiv:2112.02035 (2021).

\end{thebibliography}
\end{document}